\documentclass[preprints,article,accept,moreauthors,pdflatex]{Definitions/mdpi}  

\DeclareMathOperator\arcsinh{arcsinh}

\firstpage{1} 
\pubvolume{1}
\issuenum{1}
\articlenumber{0}
\pubyear{2024}
\copyrightyear{2024}
\datereceived{ } 
\daterevised{ } 
\dateaccepted{ } 
\datepublished{ } 
\hreflink{https://doi.org/} 

\Title{Strange quark stars: the role of excluded volume effects}

\TitleCitation{Strange quark stars: the role of excluded volume effects}

\Author{G. Lugones $^{1}$\orcidA{}, and A.G. Grunfeld $^{2,3}$\orcidB{}}

\AuthorNames{Firstname Lastname, Firstname Lastname and Firstname Lastname}

\AuthorCitation{Lugones, G.; Grunfeld, A. G.}

\address{%
$^{1}$ \quad Universidade Federal do ABC, Centro de Ci\^encias Naturais e Humanas, Avenida dos Estados 5001- Bang\'u, CEP 09210-580, Santo Andr\'e, SP, Brazil; german.lugones@ufabc.edu.br\\
$^{2}$ \quad CONICET, Godoy Cruz 2290, Buenos Aires, Argentina; ag.grunfeld@conicet.gov.ar \\
$^{3}$ \quad Departamento de F\'\i sica, Comisi\'on Nacional de Energ\'{\i}a At\'omica, Avenida del Libertador 8250, (1429) Buenos Aires, Argentina.
}

\corres{Correspondence: german.lugones@ufabc.edu.br}

\abstract{We study cold strange quark stars employing an enhanced version of the quark-mass density-dependent model which incorporates excluded volume effects to address non-perturbative QCD repulsive interactions. We provide a comparative analysis of our mass formula parametrization with previous models from the literature. 
We identify the regions within the parameter space where three-flavor quark matter is more stable than the most tightly bound atomic nucleus (stability window). Specifically, we show that excluded volume effects do not change the Gibbs free energy per baryon at zero pressure,  rendering the stability window unaffected. 
The curves of pressure versus energy density exhibit various shapes -- convex upward, concave downward, or nearly linear -- depending on the mass parametrization. This behavior results in different patterns of increase, decrease, or constancy in the speed of sound as a function of baryon number density.
We analyze the mass-radius relationship of strange quark stars, revealing a significant increase in maximum gravitational mass and a shift in the curves towards larger radii as the excluded volume effect intensifies. Excluded volume effects render our models compatible with all modern astrophysical constraints, including the properties of the recently observed low-mass compact object HESSJ1731.
} 


\begin{document}

\section{Introduction}

Strange quark matter (SQM) is a type of quark matter composed of a roughly equal mixture of up, down, and strange quarks. This matter form is theoretically unique due to its absolute stability, rendering it self-bound at the quantum level. SQM's stability arises from its lower energy per baryon compared to nuclear matter, suggesting that it could be the true ground state of hadronic matter under certain conditions.

The concept of SQM can be traced back to seminal papers by Itoh \cite{Itoh:1970uw}, Bodmer \cite{Bodmer:1971we}, and Witten \cite{Witten:1984rs} who proposed that quark matter containing strange quarks could be more stable than ordinary nuclear matter. Their hypothesis, often referred to as the 'strange matter hypothesis', laid the groundwork for subsequent research in this field. These pioneering works suggested that, under extreme conditions like those in the core of neutron stars, quarks may deconfine and form a stable mixture.

To describe the properties of SQM, various effective models have been employed. The most prominent among these is the MIT Bag model, which treats quarks as free particles inside a hypothetical 'bag', confining them with a constant energy density (the bag constant). This model effectively balances the quarks' freedom with confinement, allowing for a simplified yet insightful exploration of SQM characteristics \cite{Farhi:1984qu}. Other popular models include the Nambu–Jona–Lasinio (NJL) model, which incorporates chiral symmetry breaking but does not allow the existence of self-bound matter for the most accepted parametrizations \cite{Klevansky:1992qe,Buballa:1998pr}. In contrast, the quark-mass density-dependent model (QMDDM) offers an alternative approach, where the masses of quarks depend on the baryon number density \cite{Fowler:1981rp, Chakrabarty:1989bq, Chakrabarty:1991ui, Chakrabarty:1993db, Benvenuto:1989kc, Lugones:1995vg, Benvenuto:1998tx, Lugones:2002vd, Peng:1999gh, Wang:2000dc, Peng:2000ff, Yin:2008me, Peng:2008ta, Xia:2014zaa}. As first shown in \cite{Benvenuto:1989kc}, this model stands out for achieving the confinement property through quark-mass density-dependence, without the need for an explicit bag constant. 

Strange quark stars, if they exist, are compact objects entirely composed of SQM. These stars significantly deviate from conventional neutron stars in terms of many of their physical properties \cite{Alcock:1986hz}. Theoretical calculations show them to be smaller and denser, characterized by mass-radius relationships that set them apart from their hadronic counterparts. Another key feature distinguishing strange quark stars from neutron stars is their predicted cooling behaviors. Unlike neutron stars, which cool primarily through neutrino emission and photon radiation, strange quark stars could exhibit significantly more rapid cooling rates, a consequence of the direct Urca process being unimpeded in quark matter \cite{Page:2005fq, Weber:2004kj}.
Furthermore, strange quark stars are theorized to exhibit unique gravitational wave signatures. Their different internal structures could result in oscillation modes that, when perturbed, would emit gravitational waves with characteristic frequencies \cite{Andersson:2001ev, Flores:2013yqa,  VasquezFlores:2017uor}. These could potentially be detected by advanced gravitational wave observatories, offering a novel method to probe the existence of these hypothetical stars.

Before the discovery of pulsars with masses around $2 M_{\odot}$ \cite{Demorest:2010bx, Antoniadis:2013pzd, NANOGrav:2019jur, Riley:2021pdl, Miller:2021qha}, simple EOSs for SQM, such as the standard MIT bag model and the standard QMDDM, were consistent with the astrophysical data available at the time, specifically the observed masses and radii of compact stars. However, with the observation of pulsars with $\sim 2 M_{\odot}$ the EOSs provided by the simplest versions of those effective models were found to be too soft, meaning they could not account for the high masses of these newly discovered pulsars. 
To reconcile these models with observations, the introduction of new elements is necessary. In the case of the MIT bag model, the incorporation of color superconductivity \cite{Lugones:2002zd, Horvath:2004gn} and repulsive vector interactions \cite{Lopes2021_I,Lopes2021_II} allowed the construction of strange quark star models with masses above $2 M_{\odot}$.
Similarly, changes were made to the QMDDM to account for the observational data of the more massive pulsars. These modifications often involved changes to the mass parameterization of the quasiparticles. For instance, introducing a mass term that increases with the baryon number density allowed the model to achieve $M_\mathrm{max} > 2 M_{\odot}$ \cite{Xia:2014zaa, Li:2015ida, Issifu:2024zvq, You:2023bqx}. 

In recent work, we implemented modifications to the QMDDM, with a specific focus on the role of repulsive interactions among quarks \cite{Lugones:2023zfd}. This is achieved through the excluded volume effect, which prevents quark quasiparticles from being arbitrarily close to each other. In a dense quark matter environment, as is expected in the core of neutron stars, this effect renders the EOS substantially stiffer.  In this study, we will employ the model described in Ref. \cite{Lugones:2023zfd} to explore the properties of strange quark stars.

The paper is organized as follows.   Section \ref{sec:mass_formula} introduces the mass formula incorporating excluded volume effects, followed by a comparative examination of this parameterization against others reported in the literature. In Section \ref{sec:formalism}, we present an overview of the general aspects of the QMDDM, both with and without the consideration of excluded volume, and derive the EOS for SQM that will be used in the rest of this work. Section \ref{sec:results} is dedicated to presenting our results. In Section \ref{sec:conclusions}, we succinctly summarize our key findings and present the conclusions derived from our study.

\section{Mass formula}
\label{sec:mass_formula}

The QMDDM represents quark matter as a system of quasiparticles in such a way that the energy density and particle number densities maintain the same functional form as in a non-interacting system. This is achieved by using effective masses that incorporate interaction effects. Accordingly, the equivalent mass is modeled as comprising two components: a flavor-dependent (constant) current mass $m_{0 i}$ and an interacting term $m_{\mathrm{I}}$ that depends on the medium properties:
\begin{equation}
M_i = m_{0 i} + m_I,       \qquad  i= u, d, s.
\label{eq:mass_0}
\end{equation}
To accurately represent strongly interacting quark matter as a quasiparticle system, $m_I$ must be consistent with some key properties of QCD. This formula must reflect quark confinement and asymptotic freedom, requiring $m_I$ to fulfill the following criteria:
\begin{eqnarray}
\lim _{n_B \rightarrow 0} m_{\mathrm{I}} &=\infty, \label{eq:mass_limit_1}\\
\lim _{n_B \rightarrow \infty} m_{\mathrm{I}}& =0,
\label{eq:mass_limit_2}
\end{eqnarray}
where $n_B$ reprresents the baryon number density.

An explicit mass form, derived from these principles, has been proposed. This involves expanding the mass expression into a Laurent series with respect to the Fermi momentum $p_F$ (centered around $p_F=0$), retaining leading terms of both positive and negative exponents \cite{Peng:2004ev}:
\begin{equation}
m_{\mathrm{I}}=\frac{a_{-1}}{p_F}+a_1 p_F.
\end{equation}
Since the Fermi momentum scales with $n_B^{1/3}$, the following mass formula is derived \cite{Xia:2014zaa}:
\begin{equation}
m_I =  \frac{\mathcal{D}}{n_{\mathrm{B}}^{1 / 3}}+ \mathcal{C} n_{\mathrm{B}}^{1 / 3}.
\label{eq:mass_eq1}
\end{equation}
As discussed in \cite{Xia:2014zaa} and related references, the term that scales with $n_B^{1/3}$ is critical for the maximum gravitational mass of stellar configurations to reach a value above $2M_{\odot}$, as required by present pulsar observations. Simply using the first term of Eq. \eqref{eq:mass_eq1} proves insufficient for stellar configurations to fulfill pulsar's constraints. However, there is an issue with Eq.~\eqref{eq:mass_eq1} as the term increasing with $n_B^{1/3}$ contradicts the asymptotic freedom condition because it diverges when $n_B \rightarrow \infty$. This is not surprising since Eq.~\eqref{eq:mass_eq1}, being a series expansion,  makes sense only for low $n_B$. Despite this, Eq. \eqref{eq:mass_eq1} can still be employed because the undesirable behavior associated with this term is not apparent at the typical densities present in the cores of compact stars but becomes significant at much higher densities. This is due to the very small numerical value adopted for the parameter $\mathcal{C}$ in Eq. \eqref{eq:mass_eq1}.

{Another modification to the mass formula was adopted in Ref. \cite{Chu:2012rd} to take into account that the quark-quark effective interaction in quark matter should be isospin dependent. According to this, the equivalent quark mass should vary with isospin, an aspect not taken into account in  the previous formulas. The form of this isospin dependence remains uncertain and ideally requires determination through nonperturbative QCD calculations. Ref. \cite{Chu:2012rd} proposed the following phenomenological parameterization:
\begin{equation}
M_i = m_{0 i} + m_I - \tau_q \delta D_I n_B^\alpha e^{-\beta n_B}.
\end{equation}
Here, $D_I, \alpha$, and $\beta$ are free parameters and $\tau_q$ represents the isospin quantum number of quarks—with $\tau_q = 1$ for $u$ quarks, $\tau_q = -1$ for $d$ quarks, and $\tau_q = 0$ for $s$ quarks. The isospin asymmetry is parametrized by $\delta = 3 (n_d-n_u)/ (n_d+n_u)$, being $n_i$ the particle number density of quarks of the flavor $i$.  Including isospin dependence in the mass significantly affects the symmetry energy of quark matter and the characteristics of strange quark stars, thereby enabling these objects to satisfy the $2 M_{\odot}$ constraint  \cite{Chu:2012rd}.
}

Our approach to the mass formula takes a different path. {For simplicity, we neglect the contribution of the isospin contribution. Then,} we begin with an expression that explicitly satisfies Eqs. \eqref{eq:mass_limit_1} and \eqref{eq:mass_limit_2}:
\begin{equation}
M_i = m_{0i} + \frac{C}{n_B^{a/3}}.
\label{eq:mass1}
\end{equation}
The latter expression is 'flavor blind' because its dependence on the medium is uniform across all flavors \cite{Peng:1999gh}.
{As previously mentioned in references \cite{Lugones:2022upj} and \cite{Wen:2005uf}, the parameter $a$ correlates with the form of the potential $v(r)$ governing the quark-quark interaction, expressed as $v(r) \propto r^a$. While many research papers adopt $a=1$, in this study, we opt to treat it as a free parameter and investigate the ramifications of its variation. We will revisit this issue in the discussion of Figure \ref{fig:pressure}.}
In this formula, we will incorporate all non-perturbative and higher-order perturbative effects of QCD exclusively through the 'excluded volume' effect, rather than introducing additional terms into the mass formula. As discussed in Ref. \cite{Lugones:2023zfd}, the underlying concept is that particles, when in close proximity, effectively 'exclude' a certain region around them. This exclusion is due to repulsive interactions among the particles, stemming from a combination of non-perturbative phenomena associated with confinement, higher-order perturbative QCD interactions, and the dynamic behavior of the strong coupling constant.

In practice, the excluded volume effect is incorporated by replacing the system's volume $V$ with the available volume $\tilde{V}$, accounting for the space $b(n_B)$ occupied by each quasibaryon:
\begin{align}
\tilde{V}=V -  b(n_B) N_B, 
\label{eq:volume_nB}
\end{align}
being $N_B$ the total baryon number of the system. {When implementing this prescription, various thermodynamic quantities are altered, such as the Fermi momenta \cite{Lugones:2023zfd}. Additionally, since the quasiparticle mass depends on the baryon number density, Eq. \eqref{eq:mass1} is reformulated as:}
\begin{equation}
\tilde{M}_i      =  m_{0i}+\frac{C}{(N_B / \tilde{V})^{a/3}}.
\label{eq:mass_exl_1}
\end{equation}
Introducing the factor $q$, defined as:
\begin{equation}
q(n_B) \equiv \frac{\tilde{V}}{V} =   1- n_B b(n_B) ,
\label{eq:3fl_q_definition}
\end{equation}
the mass formula with excluded volume effects then becomes:
\begin{equation}
\tilde{M}_i  =  m_{0i}+\frac{C}{(n_B/q)^{a/3}}.
\label{eq:our_mass}
\end{equation}
This mass formula was introduced in Ref. \cite{Lugones:2023zfd} and will be used throughout the rest of this work. 
In Eq. \eqref{eq:3fl_q_definition}, the ratio $q$ must necessarily lie between 0 and 1, as the available volume can never be larger than the total system's volume. This means that the parametrization of the excluded volume per particle, $b(n_B)$, cannot be chosen arbitrarily.  Instead, it must be carefully chosen to ensure that the restriction $0 < q < 1$ is always satisfied. 
The mass parametrization we have adopted includes non-perturbative QCD effects while maintaining agreement with the asymptotic freedom constraint detailed in Eq. \eqref{eq:mass_limit_2}.

\section{General overview of the QMDDM}
\label{sec:formalism}

In this section, we present a synopsis of the QMDDM EOS as developed in Refs. \cite{Lugones:2022upj} and \cite{Lugones:2023zfd}. In Sec. \ref{sec:point_like} we summarize the EOS for point-like particles and in Sec. \ref{sec:excluded_volume} we show how to incorporate excluded volume effects in that EOS.
In Sec. \ref{sec:QMDDM_excluded_volume} we use the results of Secs. \ref{sec:point_like} and \ref{sec:excluded_volume} to construct the EOS for strange quark matter.

\subsection{The EOS for point-like particles}
\label{sec:point_like}

We will describe the system as a mixture of non-interacting quarks species with effective masses $M_i$ and free electrons. The total Helmholtz free energy depends on both the system's volume and the number of particles $\{N_i\}$, and it is the sum of the contributions from each species:
\begin{equation}
F(V, \{N_i\}) = \sum_{i=u, d, s, e}  F_i , 
\end{equation}
where, 
\begin{equation}
F_i  =
\begin{cases} g V M_i^4 \chi(x_i) & \quad (i = u, d, s) , \\ 
 g_e V m_e^4 \chi(x_e)  & \quad  (\text{electrons}) ,
\end{cases}
\label{eq:F_i_summary}
\end{equation}
with $g=6$ and $g_e =2$.  The function $\chi(x)$ is defined by: 
\begin{equation}
\chi(x)  = \frac{1}{16 \pi^2}\left[ x \sqrt{x^2 + 1} (2 x^2 + 1) -  \arcsinh(x) \right] ,
\label{eq:chi}
\end{equation}
where $x$ is the dimensionless Fermi momentum given by: 
\begin{eqnarray}
x_i  = \frac{1}{M_i} \left( \frac{6 \pi^2 n_i}{g} \right)^{1/3} & \quad (i = u, d, s) ,    \label{eq:appendix_B4} \\ 
x_e  =  \frac{1}{m_e} \left( \frac{6 \pi^2 n_e}{g_e} \right)^{1/3} & \quad  (\text{electrons}),   \label{eq:appendix_B5}
\end{eqnarray}
and $n_i = N_i/V$ is the particle number density of the $i$-species.

At $T=0$, the energy density coincides with the Helmholtz free energy per unit volume. Therefore: 

\begin{equation}
\epsilon = \sum_{i=u, d, s, e}  \epsilon_i,
\end{equation}
where:
\begin{equation}
\epsilon_i  =
\begin{cases} g M_i^4 \chi(x_i) & \quad (i = u, d, s) , \\ 
 g_e m_e^4 \chi(x_e)  & \quad (\text{electrons}) .
\end{cases}
\label{eq:epsilon_i_summary}
\end{equation}
The total pressure is obtained from $p = -\left.  {\partial F} /  {\partial V}\right|_{\left\{N_j\right\}}$, and is the sum of the partial pressures of each species:
\begin{equation}
p = \sum_{i=u, d, s, e}  p_i, 
\label{eq:total_pressure_pl}
\end{equation}
being
\begin{equation}
p_i  =
\begin{cases} g M_i^4 \phi(x_i) - B_i  & \quad (i = u, d, s) , \\ 
  g_e m_e^4 \phi(x_e)  & \quad (\text{electrons}) ,
\end{cases}
\label{eq:p_i_summary}
\end{equation}
where $\phi(x)$ is defined by:
\begin{equation} 
\phi(x)  =   \frac{  x \sqrt{x^{2}+1}(2 x^{2} -3) +3 \operatorname{arcsinh}(x)  }{48 \pi^{2}} .   \label{eq:appendix_phi}
\end{equation}
The 'bag constant' $B_i$ is given by: 
\begin{equation}
\begin{aligned}
- B_i  =   g  \beta(x_i)   M_i^3 \sum_j  n_j  \frac{\partial M_i}{\partial n_j}  ,
\label{eq:bag_summary}
\end{aligned}
\end{equation}
with $\beta(x)$ defined by:
\begin{equation} 
\beta(x)  =  \frac{1}{4 \pi^{2}}\left[x \sqrt{x^{2}+1} -\operatorname{arcsinh}(x)\right] \label{eq:appendix_beta} .
\end{equation}
In the flavor blind model the mass has the form $M_i = m_{0i} + m_I(n_B)$, i.e. the interaction term $m_I$ of the mass depends only on $n_B$. Therefore:
\begin{equation}
\frac{\partial M_i}{\partial n_j} =\frac{d M_i}{d n_B} \cdot \frac{d n_B}{d n_j}=\frac{1}{3} \frac{d M_i}{d n_B} = \frac{d M_i}{d n_i}.
\end{equation}
Thus, for flavor blind mass formulas,  Eq. \eqref{eq:bag_summary} reads:
\begin{equation}
-B_i=g \beta\left(x_i\right) M_i^3  \frac{\partial M_i}{\partial n_i} \sum_j n_j = g \beta\left(x_i\right) M_i^3  \frac{\partial M_i}{\partial n_i} 3 n_B = g \beta\left(x_i\right) M_i^3  n_B \frac{\partial M_i}{\partial n_B} .  
\label{eq:bag_blind}
\end{equation}
In our previous publication \cite{Lugones:2022upj}, an error was identified in the definition of the bag constant which was further corrected in  \cite{Lugones:2023zfd}. Eq. \eqref{eq:bag_summary}  represents the accurate expression for any mass formula, while Eq. \eqref{eq:bag_blind} applies specifically to the flavor blind scenario.
By substituting Eq. \eqref{eq:bag_summary} into \eqref{eq:p_i_summary}, and subsequently replacing  Eq. \eqref{eq:p_i_summary} into \eqref{eq:total_pressure_pl}, we arrive at the same expression for the pressure as given in Refs. \cite{Peng:2008ta, Xia:2014zaa}.

Finally, the chemical potential of the $i$-species is derived from $\mu_i  ={\partial F} / {\partial N_i}|_{V,\left\{N_{j \neq i}\right\}}$. For any mass formula, the chemical potential of $u$, $d$, and $s$ quasiparticles is given by the following expression (cf. Appendix B of \cite{Lugones:2023zfd}):
\begin{equation}
\mu_i  =   M_i \sqrt{x_i^{2}+1} + \sum_j g M_j^3  \frac{\partial M_j}{\partial n_i} \beta(x_j) ,
\label{eq:mu_summary}
\end{equation}
which coincides with the expressions given in Refs. \cite{Peng:2008ta, Xia:2014zaa}.
In the flavor blind scenario, the latter equation simplifies to:
\begin{equation}
\mu_i  = M_i \sqrt{x_i^{2}+1} -   \frac{1}{3 n_B} \sum_{j}  B_j .
\label{eq:mu_summary2}
\end{equation}
For the electron component, the chemical potential is given by:
\begin{eqnarray}
\mu_e = m_e \sqrt{x_e^{2}+1} . 
\label{eq:mu_summary_2}
\end{eqnarray}
It should be noted that the second term of Eq. \eqref{eq:mu_summary2} differs from the expression in Eq. (88) of Ref. \cite{Lugones:2022upj}. The version presented above is indeed the accurate one.

\subsection{Excluded volume effects: general formulas}
\label{sec:excluded_volume}

As shown in Ref. \cite{Lugones:2023zfd}, the incorporation of excluded volume effects into any zero-temperature EOS originally formulated for point-like particles in the Helmholtz representation is straightforward. The procedure unfolds as follows. Initially, we select an ansatz for the excluded volume per particle, $b(n_B)$, from which we derive the  following quantities:
\begin{eqnarray} 
q &=& 1 - n_B b ,  \label{eq:summary_q_3fl} \\
\delta &=& 1 + n_B^{2} \frac{db}{dn_B} ,  \label{eq:summary_delta_3fl}  \\
\lambda &=& \frac{1}{3} \left( b + n_B \frac{db}{dn_B} \right) . \label{eq:summary_lambda_3fl}    
\end{eqnarray}
It's worth noting that the functional form of $b(n_B)$ must be selected to ensure that the constraint $0 < q < 1$ is always verified.

Subsequently, we adapt the expressions for the energy density $\epsilon_{\mathrm{pl}}$, pressure $p_{\mathrm{pl}}$, and chemical potentials $\mu_{\mathrm{pl}, i}$ of point-like particles. This involves rewriting $\epsilon_{\mathrm{pl}}$, $p_{\mathrm{pl}}$, and $\mu_{\mathrm{pl}, i}$ in terms of the modified variable set $\{n_j/q\}$, and then multiplying them by the  factors $q$, $\delta$, and $\lambda$ as outlined below:
\begin{eqnarray}
\epsilon(\{n_j \}) & = & q \sum_{i}  \epsilon_{\mathrm{pl},i}    ( \left\{ n_j /q  \right\} )  , \label{eq:summary_epsilon_3fl} \\
p( \left\{ n_j \right\} ) & = &   \delta  \sum_{i}   p_{\mathrm{pl},i} ( \left\{ n_j /q  \right\} ) , \label{eq:summary_p_3fl}\\
\mu_k( \{n_j\} ) & = & \lambda(n_B) p_{\mathrm{pl}} \left(  \left\{ \tfrac{n_j}{q}  \right\}   \right)  + \mu_{\mathrm{pl}, k}  \left(  \left\{ \tfrac{n_j}{q}  \right\}   \right) . \qquad  \label{eq:summary_mu_3fl}
\end{eqnarray}
In this manner, we obtain expressions that incorporate the effect of excluded volume, derived from the known expressions for point-like particles.

\subsection{Strange quark matter EOS}
\label{sec:QMDDM_excluded_volume}

Now, we adopt an ansatz for $b$ consistent with the asymptotic freedom behavior of QCD, i.e., 
we adopt a formula that allows $b$ to approach zero at asymptotically large densities.  Based on the prescription of Ref. \cite{Lugones:2023zfd}, the excluded volume is defined as
\begin{equation}
b = \frac{\kappa}{n_B} ,
\label{eq:ansatz_figures}
\end{equation}
being $\kappa$ a positive constant. 
Using Eqs. \eqref{eq:summary_q_3fl}, \eqref{eq:summary_delta_3fl} and \eqref{eq:summary_lambda_3fl},  the functions $q$, $\delta$ and $\lambda$ are:
\begin{eqnarray}
q(n_B) &=&   1 - \kappa, \\
\delta(n_B) &=&  1 - \kappa, \\
\lambda(n_B) &= & 0 .
\end{eqnarray}
Replacing these parameters in Eqs. \eqref{eq:summary_epsilon_3fl}, \eqref{eq:summary_p_3fl} and \eqref{eq:summary_mu_3fl} we obtain:
\begin{eqnarray}
\epsilon &=& \sum_{i=u, d, s}  (1 - \kappa) g  \tilde{M}_i^4 \chi(\tilde{x}_i)  + \epsilon_e ,   \label{eq:epsilon_figures} \\
p &=& \sum_{i=u, d, s}  (1 - \kappa)  (  g  \tilde{M}_i^4 \phi(\tilde{x}_i) - \tilde{B}_i)     + p_e ,  \label{eq:pressure_figures} \\
\mu_i   &=&  \tilde{M}_{i} \sqrt{\tilde{x}_{i}^{2}+1}-    \frac{1}{3 \tilde{n}_B}    \sum_{j}     \tilde{B}_{j}    , 
\label{eq:chemical_figures} 
\end{eqnarray}
where
\begin{eqnarray}
\tilde{n}_B &=& n_B/q,  \\
\tilde{M}_i &=& M_i(\tilde{n}_B),  \\
\tilde{x}_i  &=& \frac{1}{\tilde{M}_i} \left[ \frac{6 \pi^2 (n_i/q) }{g} \right]^{1/3}  \qquad (i = u, d, s), \\   
-\tilde{B}_i &=& g \beta\left(\tilde{x}_i\right) \tilde{M}_i^3  \tilde{n}_B \frac{\partial \tilde{M}_i}{\partial \tilde{n}_B} . 
\end{eqnarray}

As our focus is on describing quark matter within the interiors of cold compact stars, we will enforce the conditions of local electric charge neutrality and chemical equilibrium under weak interactions. Under these conditions, neutrinos are free to exit the system ($\mu_{\nu_{e}}=0$), leading to the following chemical equilibrium conditions:
\begin{eqnarray}
\mu_{d} &=& \mu_{u}+\mu_{e}, \\
\mu_{s} &=& \mu_{d} .   
\end{eqnarray}
Meanwhile, charge neutrality can be expressed as:
\begin{equation}
\tfrac{2}{3} n_{u}-\tfrac{1}{3} n_{d}-\tfrac{1}{3} n_{s}-n_{e}=0 .    
\end{equation}

\section{Numerical results for the EOS under compact star conditions}
\label{sec:results}

In this section, we will use the EOS introduced in Section \ref{sec:QMDDM_excluded_volume} to explore the properties of strange quark matter and strange quark stars. {We will explore various values for the parameters $a$, $C$, and $\kappa$ of the model. The parameter $\kappa$ is dimensionless. The units of $a$ and $C$ are not fixed and are chosen such that $n_B$ is in $\mathrm{fm}^{-3}$ and $M_i$ is in $\mathrm{MeV}$ in Eq. \eqref{eq:our_mass}.}

\begin{figure}[tb]
\centering
\includegraphics[width=0.99\columnwidth,angle=0]{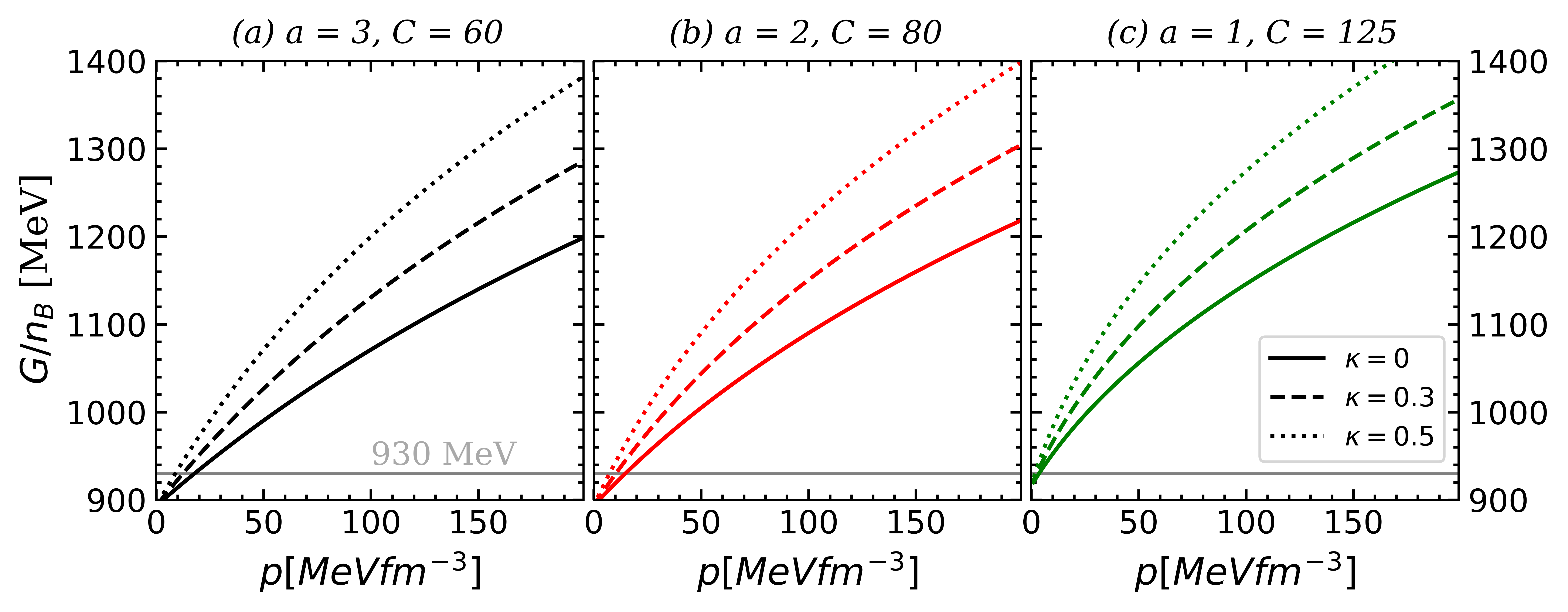}
\caption{Gibbs free energy per baryon $G/n_B$ as a function of pressure $p$ for different parameter sets labeled (a), (b), and (c). Within each panel, three values of the excluded volume parameter $\kappa$ are considered: solid lines represent point-like particles ($\kappa = 0$), dashed lines correspond to $\kappa = 0.3$, and dotted lines to $\kappa = 0.5$. The horizontal line at $930$ MeV indicates the energy per baryon in $^{62}\mathrm{Ni}$, serving as a benchmark for stability against decay into normal nuclear matter. Notice that within each panel all curves coincide at $p=0$ (see discussion in the text for further details on this issue). }
\label{fig:gibbs}
\end{figure}

Figure \ref{fig:gibbs} illustrates the Gibbs free energy per baryon,  $G / n_{B} = (\epsilon+p) / n_{B}$, plotted against pressure. Depending on the chosen parameters $a$, $C$, and $\kappa$, the value of $G / n_{B}$ at zero pressure may either be above or below the energy per nucleon in the most tightly bound atomic nucleus, $^{62}\mathrm{Ni}$, which is approximately 930 MeV.
For parameter sets where $G / n_{B}$ is less than 930 MeV at zero pressure and temperature, we are in the case of self-bound quark matter. In these cases, bulk quark matter is stable in vacuum and does not convert into hadronic matter. If this self-bound matter includes all three flavors ($u$, $d$, and $s$) and leptons at $p=T=0$, it is known as SQM. In such circumstances, it is conceivable for compact stars composed entirely of quark matter, termed self-bound quark stars, to exist.
On the other hand, if $G / n_{B}$ exceeds 930 MeV, the matter is classified as hybrid matter, which is characterized by a transition from a hadronic state at lower pressures to a deconfined state at higher pressures. In such cases, stars containing quark matter would be hybrid stars, featuring a core of quark matter surrounded by hadronic matter.

In Figure \ref{fig:gibbs}, we selected parameter sets to ensure that the $G / n_{B}$ curves for $uds$ quark matter are positioned below 930 MeV at zero temperature and pressure. It is important to note that these $G / n_{B}$ curves are sensitive to variations in the parameter $\kappa$. Despite this, the value of $G / n_{B}$ at zero pressure—a critical factor in determining the self-bound nature of quark matter—is unaffected by changes in $\kappa$. 
This is evident in the three panels of Figure \ref{fig:gibbs}, where the curves converge at the same point within each panel.

\begin{figure}[tb]
\centering
\includegraphics[width=0.7\columnwidth,angle=0]{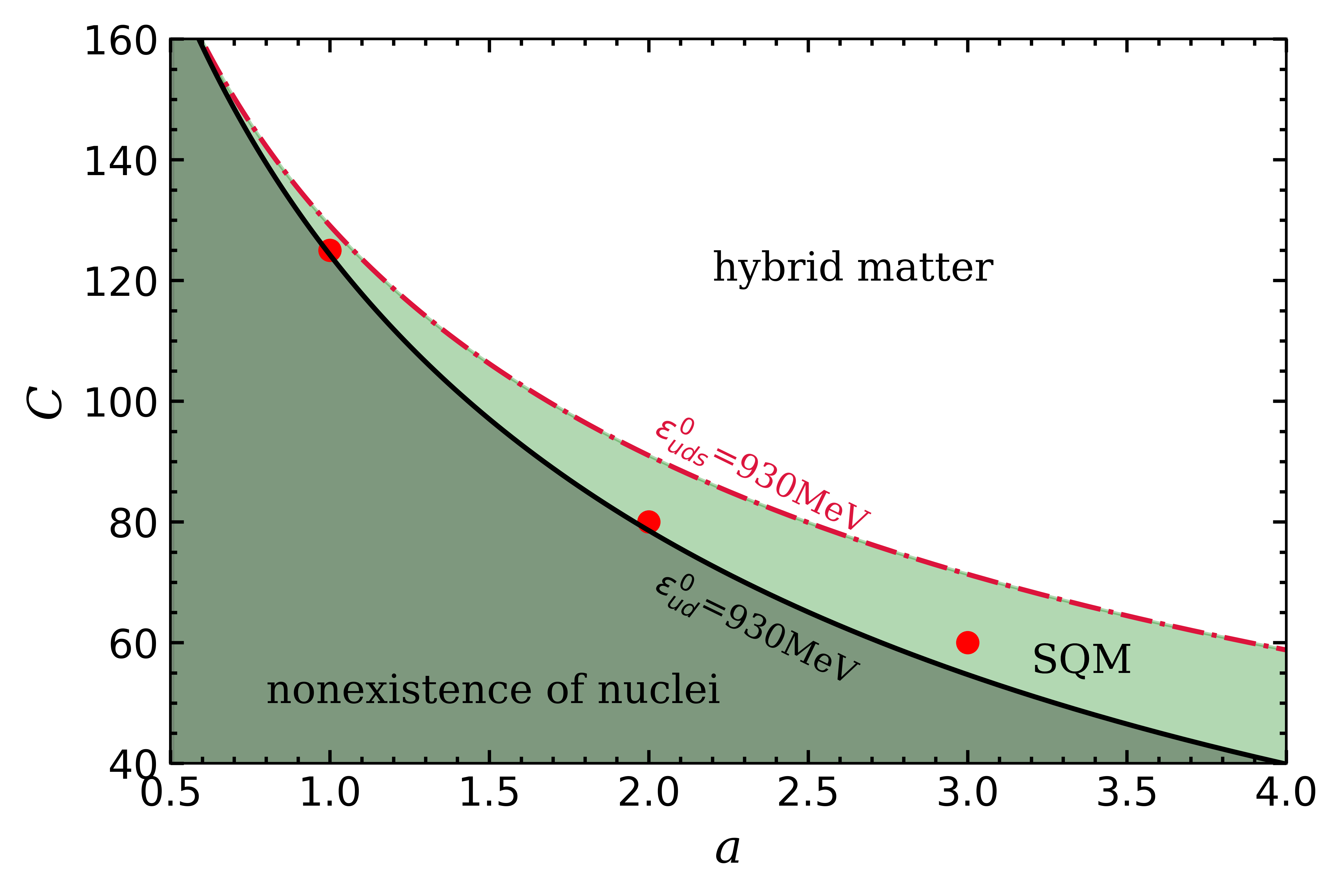}
\caption{Parameter space for quark matter in bulk. The shaded light green region denotes the parameter combinations for which SQM is the true ground state of matter. The region above the light green area, labeled as 'hybrid matter', corresponds to a scenario where hadronic matter is energetically favored at low pressures, whereas quark matter becomes preferred at higher pressures. The dark green region below, marked as 'nonexistence of nuclei', represents parameter choices where nuclei would be unstable. The same parameter space is valid for any choice of the excluded volume parameter $\kappa$. Note that the stability window presented here has some quantitative differences from that shown in  \cite{Lugones:2022upj}. The reason for this discrepancy is that Ref.~\cite{Lugones:2022upj} contained an error in the expression of the chemical potential, as previously mentioned at the end of Section \ref{sec:point_like}.}
\label{fig:window}
\end{figure} 

Figure \ref{fig:window} presents the stability window of quark matter as determined by the parameters $a$ and $C$. As previously explained, the value of the Gibbs free energy per baryon at zero pressure does not depend on the value of $\kappa$. Therefore, the stability window shown in the figure holds for any value of the excluded volume. 
The plot includes two significant curves. The solid black curve labeled with $e_{ud}^0 = 930 \mathrm{MeV}$ represents parameter values at which $G / n_{B}$ at zero pressure is exactly 930 MeV for two-flavor quark matter. The dashed red curve with $e_{uds}^0 = 930 \mathrm{MeV}$ corresponds to parameter values where $G / n_{B}$ at $p=0$ is exactly 930 MeV for three-flavor quark matter.
The light green shaded area in between indicates the parameter values for $a$ and $C$ where three-flavor quark matter is more stable than the most tightly bound atomic nucleus, and two-flavor quark matter is less stable than this nucleus (strange matter hypothesis). The EOS models represented by these parameters support the possibility of self-bound quark matter without contradicting the existence of atomic nuclei.
Within this region, we have selected three sets of parameters that will be used for the calculations in the following figures, which are identified by red dots in the stability window.
The dark green region corresponds to parameter values where two-flavor quark matter (in bulk) is more stable than the most tightly bound atomic nucleus. This area represents a scenario where atomic nuclei could decay into a single bag of deconfined $u$ and $d$ quarks, which is not observed in Nature. 
Additionally,  it is important to note that throughout the stability window, the energy per baryon for matter with three flavors always remains lower than for two-flavor matter, meaning that the present model does not support the existence of absolutely stable two-flavor matter. This means that in the current model, the dark green region represents a scenario where nuclei would ultimately decay into small bags of quark matter consisting of $u$, $d$, and $s$ quarks. One might speculate that the parameters of the dark green region are physically viable because the existence of such small hypothetical configurations might be practically suppressed by large surface and curvature tensions. In such a scenario, self-bound quark matter would be possible in bulk, but suppressed within nuclei. 
Yet, this paper opts for a more restrictive approach, considering the dark green area as unphysical.
In the white region, the parameters $a$ and $C$ render both two- and three-flavor quark matter less stable than the most tightly bound atomic nucleus. Here, quark matter could only exist within the core of compact stars, leading to the formation of what are known as hybrid stars. 
The range of $a$ values depicted ensures that the speed of sound is always lower than the speed of light throughout the figure.

\begin{figure}[tb]
\centering
\includegraphics[width=0.99\columnwidth,angle=0]{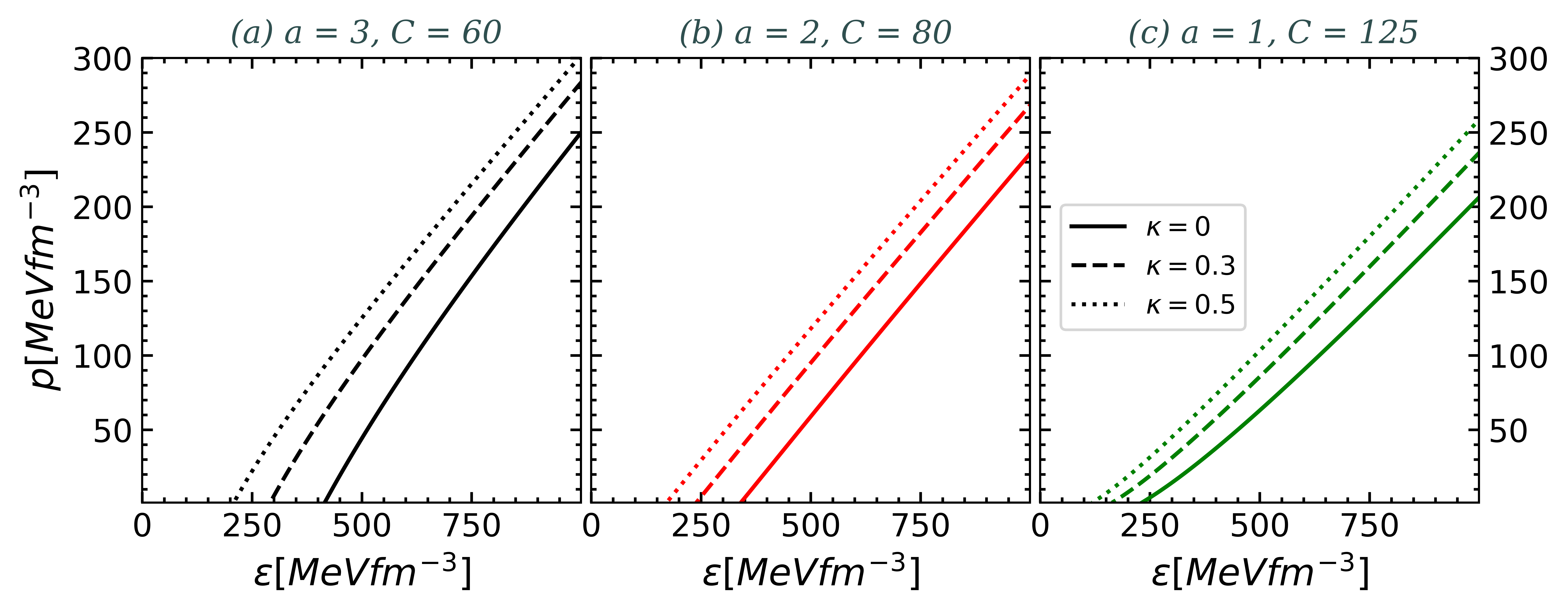}
\caption{Pressure $p$ as a function of the energy density $\epsilon$ for the same sets of model parameters as shown in Figure \ref{fig:gibbs}. In panel (a), the curves exhibit concave downward behavior; in panel (b), they are straight lines; and in panel (c), they are concave upward. As the value of $\kappa$ increases, the EOS becomes stiffer.}
\label{fig:pressure}
\end{figure}

In Figure \ref{fig:pressure}, we present the total pressure $p$ as a function of the energy density $\epsilon$ using the same parameter choices as in Figure \ref{fig:gibbs}. In each case, the pressure turns negative at a finite energy density due to the effect of the bag constant $B$. The EOS is highly sensitive to the parameter $\kappa$, with an increase in $\kappa$ consistently leading to a stiffer EOS. As $\kappa$ increases, the bag constant diminishes, causing the point at which the pressure of matter vanishes to move to lower energy densities. Although not apparent in the figure due to the chosen scale, the excluded volume effect disappears at high densities, which results in the convergence of the curves for different $\kappa$ values.

An interesting feature of the parameterizations used in each panel is the change in the concavity of the curves. In panel (a), the curves are concave downwards, in panel (b) they are almost linear, and in panel (c) they are concave upwards.
As discussed in Refs. \cite{Lugones:2022upj} and \cite{Wen:2005uf}, the parameter $a$ is related to the form of the potential $v(r)$ of the quark-quark interaction, such that $v(r) \propto r^a$ (this parameter $a$  appears in the mass as $m_{\text{I}} \propto n_B^{-a / 3}$). Accordingly, panel (a) corresponds to a cubic potential, panel (b) to a quadratic potential, and panel (c) to a linear potential.
It is important to note that the change in concavity of the EOS occurs for quadratic interactions. Most research papers use $a=1$, corresponding to a linear interaction potential.

\begin{figure}[tb]
\centering
\includegraphics[width=0.99\columnwidth,angle=0]{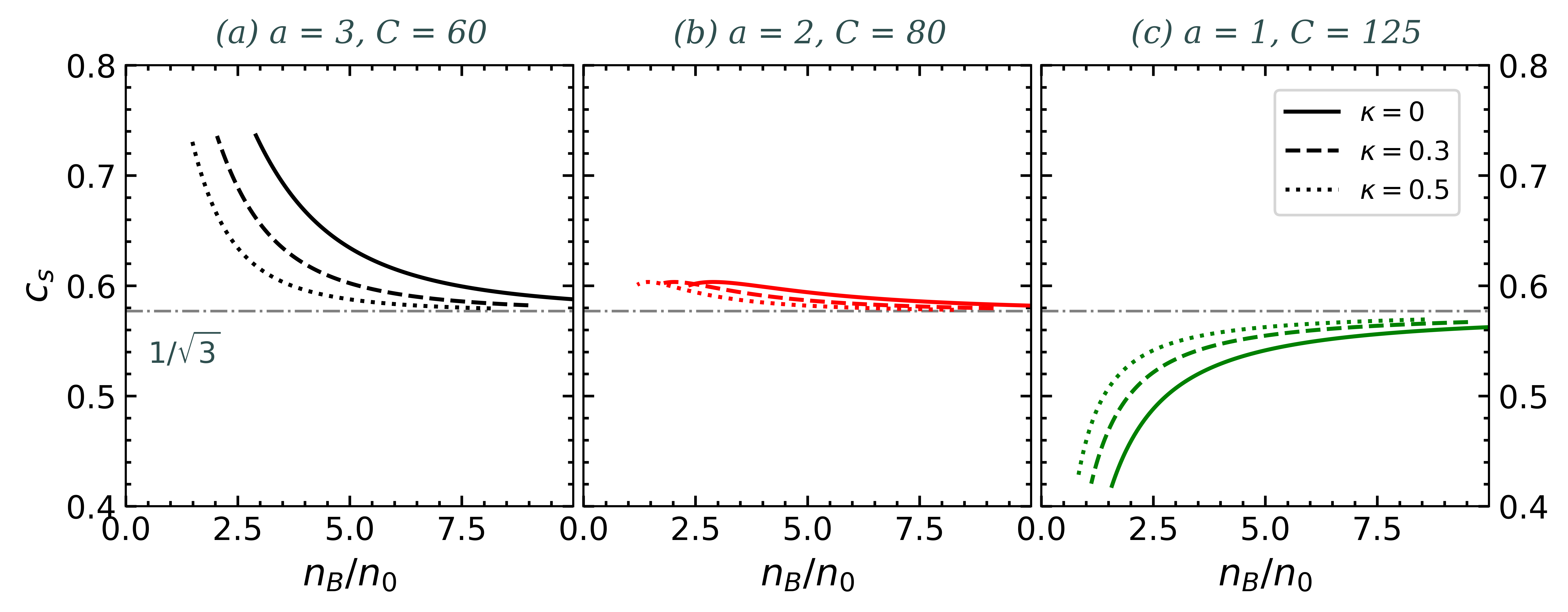}
\caption{Speed of sound $c_s$, normalized to the speed of light, as a function of the baryon number density normalized to nuclear saturation density $n_B/n_0$. In panel (a) the curves show a decreasing trend which corresponds to the concave curvature seen in the pressure versus energy density plot in the previous figure. Panel (b) presents nearly horizontal curves, consistent with the near-linear relation between $p$ and $\epsilon$ observed previously. Finally, in panel (c) the curves are ascending reflecting the convex curvature of the curves in the previous figure.  The gray horizontal line indicates the conformal limit of $c_s = 1/\sqrt{3}$. It is observed that $c_s$ approaches this limit at high densities.   }
\label{fig:sound}
\end{figure}

The speed of sound $c_{s}$ is shown in Figure \ref{fig:sound} for the same parameter choices of previous figures.  In every scenario, $c_{s}$ approaches the conformal limit, where $c_{s}=1 / \sqrt{3}$ asymptotically. Since the pressure approaches zero at a finite density, we truncate the curves at the density where $p = 0$.  Notice that for $a=3$, $c_s$ decreases with increasing $n_B$; for $a=2$, it remains nearly constant; and for $a=1$, it increases with $n_B$. We have numerically confirmed that this change in behavior is directly associated with the parameter $a$ (linked to the quark-quark interaction potential), while the parameter $C$ plays a minor role in this feature. This transition in behavior is observed very close to $a=2$.
When $c_{s}$ is a decreasing function of density (panel a), we observe that as the value of the parameter $\kappa$ grows,  $c_s$ is smaller for a given density. Conversely, when $c_s$ rises with density, an increase in the value of $\kappa$ leads to a higher speed of sound at a given density. These differences disappear at asymptotically high densities, where all curves ultimately converge.

\begin{figure}[tb]
\centering
\includegraphics[width=0.99\columnwidth,angle=0]{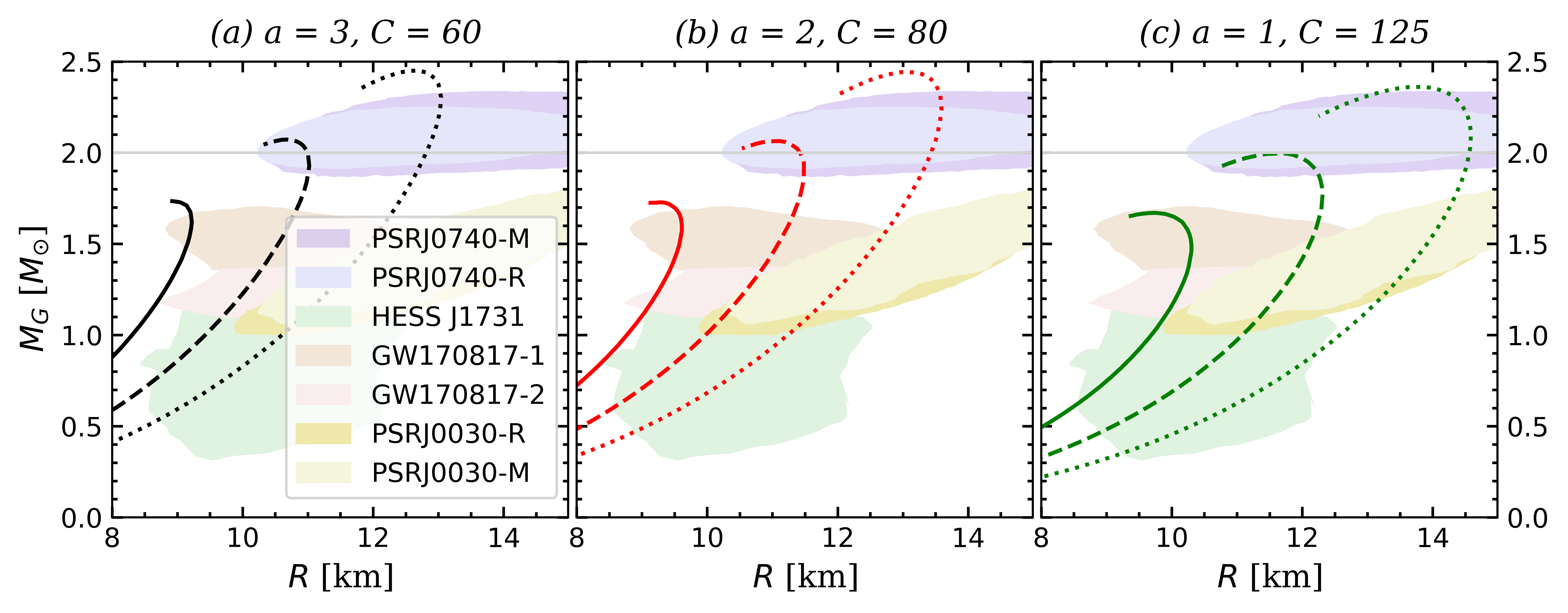}
\caption{Mass-radius relationship for strange quark stars calculated under the same parameter sets of previous figures.  The solid, dashed, and dotted lines within each panel have the same meaning as in previous figures. The different shaded regions correspond to observational data from various sources, providing constraints on the mass and radius of observed NSs.}
\label{fig:stars}
\end{figure}

In Figure \ref{fig:stars}, we present the gravitational mass $M_G$ versus radius $R$ relationship for strange quark stars, using the same parameter sets as in the previous figures. The following characteristics are consistent across all panels of Figure \ref{fig:stars}. For point-like quarks ($\kappa = 0$), the maximum $M_G$ does not satisfy the $2 M_{\odot}$ constraint, and the mass-radius curve does not match the astrophysical constraints established by NICER \cite{Riley:2019yda,Miller:2019cac,Riley:2021pdl,Miller:2021qha} observations. 
When taking into account the excluded volume of quasiparticles, the EOS becomes stiffer, as previously mentioned. {This leads to an increased maximum gravitational mass and larger stellar radii at a given mass, and the resulting curves satisfy the main astrophysical constraints.}
Notably, the maximum gravitational mass can reach nearly $2.5 M_{\odot}$ when $\kappa = 0.5$, indicating a significant rise due to the effects of the excluded volume.
{Finally, the curves with the excluded volume effect shown in Figure \ref{fig:stars} also satisfy the constraints imposed by the observation of HESSJ1731 \cite{HESS2022}. This object has both a very small mass and radius, making it difficult to describe using hadronic matter models.}

\begin{figure}[tb]
\centering
\includegraphics[width=0.99\columnwidth,angle=0]{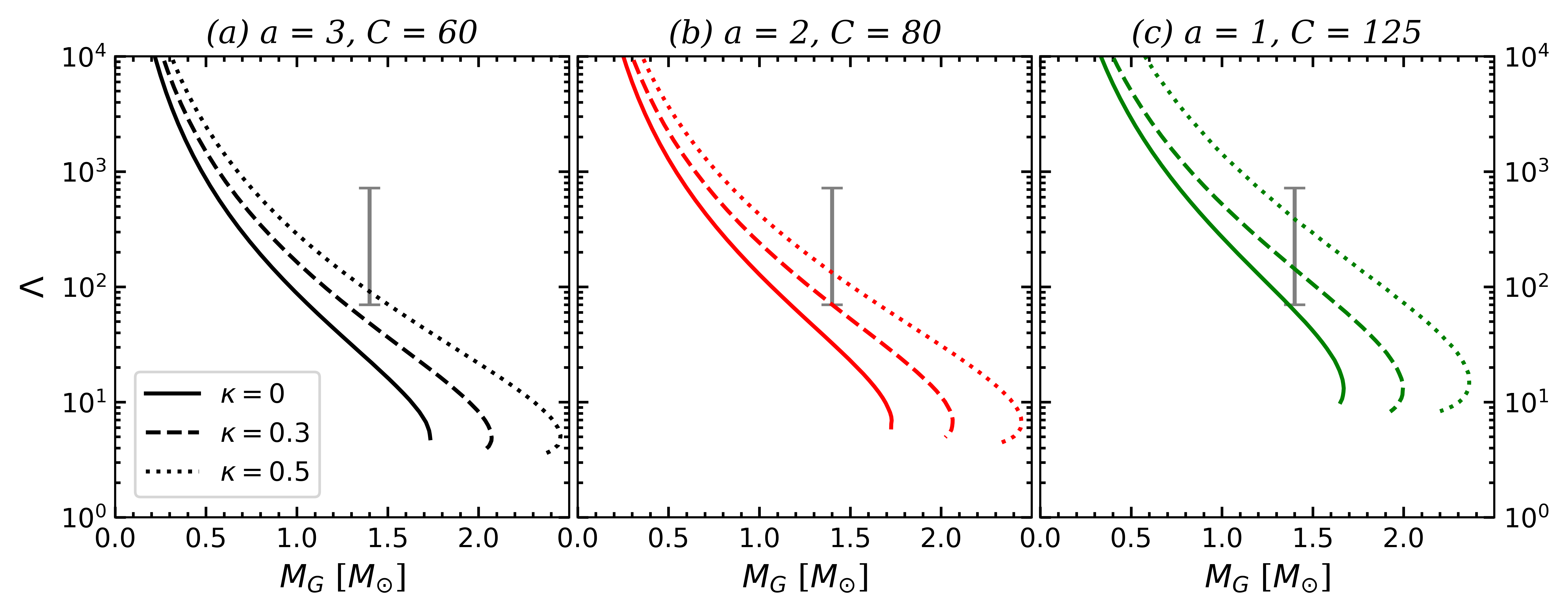}
\caption{{Variation of tidal deformability ($\Lambda$) with the star gravitational mass ($M_G$) for the same model parameters as in previous figures. The error bar is the tidal deformability of a 1.4 $M_{\odot}$ neutron star estimated from GW170817 to be $\Lambda_{1.4}=$ $190_{-120}^{+390}$ at the $90 \%$ level. }}
\label{fig:tidal_def}
\end{figure}

{
In Figure  \ref{fig:tidal_def} we show the behavior of the tidal deformability $\Lambda$ as a function of strange star gravitational mass $M_G$ across different sets of model parameters. Neutron stars experience significant tidal deformation during the inspiral phase of a neutron star-neutron star merger leaving a detectable imprint on the observed gravitational waveform of the merger.  The extent of this effect is expressed in terms of the star's dimensionless tidal deformability,  which characterizes the induced mass quadrupole moment in response to an external tidal field.  Specifically, the induced mass-quadrupole moment $Q_{ij}$ and the applied tidal field $\epsilon_{ij}$ are related to this parameter by $Q_{ij} = -\Lambda M_G^5 \epsilon_{ij}$.  The dimensionless tidal deformability parameter can be expressed as:
\begin{equation}
\Lambda = \frac{2}{3} k_2 \mathbb{C}^{-5},
\label{eq:tidal_deformability}
\end{equation}
where $\mathbb{C} \equiv M_G/R$ is the compactness of the star and $k_2$ is the tidal Love number \cite{Postnikov:2010yn}.  
In the three panels of Figure \ref{fig:tidal_def}, a general feature observed is the steep decrease in $\Lambda$ with increasing $M_G$. Alongside this trend, as the parameter $\kappa$ increases, there is an intensification of repulsive forces within strange quark matter, leading to greater tidal deformability at a fixed  $M_G$. 
The error bar represents the estimated tidal deformability of a 1.4 $M_{\odot}$ neutron star from GW170817, which is $\Lambda_{1.4} = 190_{-120}^{+390}$ at the $90\%$ confidence level. Notably, the model with $a=1$ aligns more closely with the GW170817 error bar, particularly when compared to the model with $a=3$, where compatibility is only observed for $\kappa=0.5$. This difference can be understood through the impact of the parameter $\mathbb{C}$ on tidal deformability. As seen in  Figure \ref{fig:stars}, decreasing $a$ shifts the curves toward regions of larger radii, reducing compactness and enhancing tidal deformability. 
}

\begin{figure}[tb]
\centering
\includegraphics[width=0.99\columnwidth,angle=0]{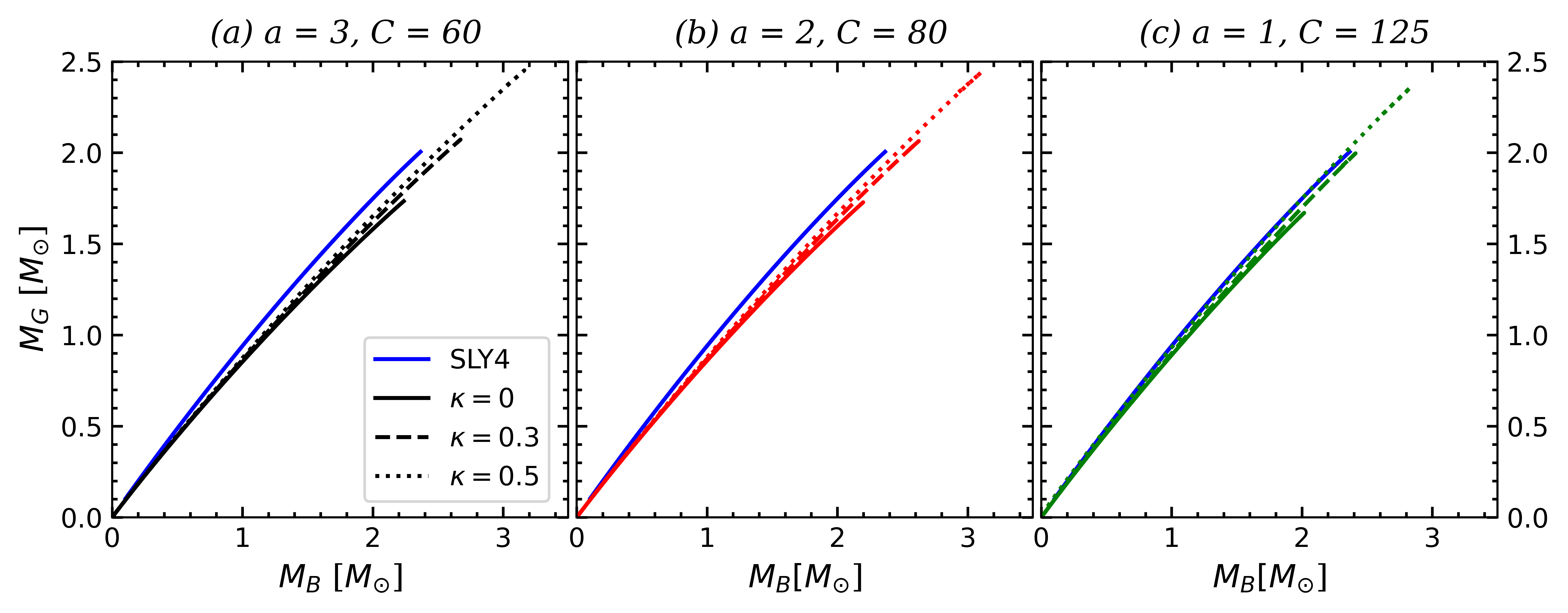}
\caption{{Relationship between gravitational mass ($M_G$) and  baryon mass ($M_B$)  for the same model parameters as in previous figures. For comparative purposes, we have included the curve representing the SLY4 hadronic EOS. Each panel illustrates how $M_G$ increases roughly linearly with $M_B$ across different model parameters.}}
\label{fig:baryon_mass}
\end{figure}

{
Finally, Figure \ref{fig:baryon_mass} displays the gravitational mass of the star as a function of its baryon mass ($M_B$). The baryon number for a canonical neutron star ($M_G=1.4 M_{\odot}$) can be estimated as $A_{\mathrm{B}} \simeq 1.4 M_{\odot} / m_n = 1.7 \times 10^{57}$. It is useful to define the equivalent mass of non-interacting baryons, termed the baryon mass (or rest mass) of the star, denoted as $M_{\mathrm{B}} \equiv A_{\mathrm{B}} m_{\mathrm{B}}$, where $m_{\mathrm{B}} \approx m_n$ represents the mass of one baryon. A common choice for $m_{\mathrm{B}}$ is to set it equal to the neutron mass, $m_n$. This assumption leads to the equation:
\begin{equation}
M_{\mathrm{B}} = A_{\mathrm{B}} m_n \simeq 0.842 A_{\mathrm{B} 57} M_{\odot},
\end{equation}
where $A_{\mathrm{B} 57} \equiv A_{\mathrm{B}} / 10^{57}$. 
The values of $A_{\mathrm{B}}$ and $M_{\mathrm{B}}$ remain constant throughout the evolutionary path of an isolated star. This constancy in baryon mass is essential for understanding the potential transformation of a metastable hadronic star into a strange quark star. In this proposed scenario, it is hypothesized that the strange star does not form directly during a supernova explosion but rather emerges through a phase conversion from another neutron star. During such a transformation, the baryon mass must remain unchanged between the initial and final states. To illustrate this, we have included in Figure \ref{fig:baryon_mass} the curve for the widely used SLY4 hadronic EOS \cite{Douchin:2001sv} for comparison. As previously mentioned, the phase conversion of a hadronic star occurs along a vertical trajectory in Figure \ref{fig:baryon_mass}. A critical aspect of this figure is that the SLY4 curve is consistently above those of the strange matter objects, with few exceptions. This positioning indicates that the conversion of a hadronic star into a strange star is an exothermic process. This finding qualitatively aligns with the results obtained using different EOS for strange quark matter \cite{Berezhiani:2002ks, Bombaci:2004mt, Drago:2004vu}.
}

{
In all panels of Figure \ref{fig:baryon_mass}, the curves demonstrate that as $\kappa$ increases, the slope of the curve becomes steeper. This suggests that a higher $\kappa$ value, indicative of stronger repulsive interactions within the neutron star, leads to a higher gravitational mass for the same baryon mass.
Generally, the curves in panel (a) exhibit a steeper slope than those in panel (b), which in turn are steeper than those in panel (c). This pattern is associated with the fact that stiffer EOSs (see Figure \ref{fig:pressure}) result in a greater gravitational mass for a fixed baryon mass. Consequently, as the parameter $a$ decreases, the energy released decreases, as evidenced by the narrowing gap between the SLY4 and the SQM curves.
}

\section{Summary and Conclusions}
\label{sec:conclusions}

In this work, we examined the properties of strange quark matter using the recently developed QMDDM with excluded volume effects, as detailed in \cite{Lugones:2023zfd}. Our study focuses on the properties of self-bound compact stars formed by this matter.

High-density quark matter at zero temperature is represented as a system of quasiparticles whose effective masses depend on the baryon number density. In the initial version of the model \cite{Lugones:2022upj, Peng:2008ta, Xia:2014zaa}, the mass formula for these quasiparticles phenomenologically incorporates key quark properties such as color confinement, asymptotic freedom, and chiral symmetry breaking/restoration. This is achieved through an effective mass that diminishes with increasing $n_B$. A recent improvement in this model accounts for potential repulsive interactions between quasiparticles by considering them as surrounded by an 'excluded volume'.  In our current study, we contrasted our mass formula with those proposed in earlier studies, particularly those that incorporate an additional term that grows with $n_B$ \cite{Xia:2014zaa}. It is argued that this additional term represents leading-order perturbative interactions in QCD. 
However, such an additional term contradicts the property of asymptotic freedom which requires that the mass of quasiparticles tend to the current quark mass at high $n_B$. In practice, this inconsistency 
does not pose significant issues for astrophysical applications, as the EOS is not applied in regimes of extremely high density.  However, in a more consistent approach it is crucial to guarantee that the effects of any interaction vanish at asymptotically high densities. Thus, this article focused on a mass formula incorporating an excluded volume that decreases with $n_B$ and vanishes as $n_B \rightarrow \infty$.

Our results show that the Gibbs free energy per baryon at zero pressure remains unaffected by the $\kappa$ parameter associated with the excluded volume per quasiparticle. Consequently, the stability window shown in Figure \ref{fig:window} is the same across different $\kappa$ values. Within this stability window, we focused on the light green shaded area, associated with parameter values for $a$ and $C$ where three-flavor quark matter is more stable than the most tightly bound atomic nucleus, while two-flavor quark matter falls short of this stability (commonly referred to as the SQM hypothesis). The EOS models represented by these parameters support the possibility of strange quark stars.

From the SQM region, we selected three parameter sets for our calculations, indicated by red dots in Figure \ref{fig:window}.
The pressure curves in Figure \ref{fig:pressure} exhibit a strong dependence on the parameter $a$, which is manifested in variations of concavity. Specifically, our numerical findings indicate that the curves are convex upward for $a \lesssim 2$, linear for $a \approx 2$, and concave downward for $a \gtrsim 2$. Additionally, an increase in the $\kappa$ parameter leads to an upward shift of the entire curve, i.e., at any given energy density, a higher $\kappa$ value corresponds to increased pressure. Such a behavior indicates a rise in the stiffness of the EOS for larger $\kappa$ values.
Correspondingly, the shape of the pressure-energy density curves directly influences the behavior of the speed of sound as a function of baryon number density, as apparent from Figure \ref{fig:sound}. For $a \lesssim 2$, $c_s$ increases with $n_B$; for $a \approx 2$, the curves are almost horizontal; and for $a \gtrsim 2$, $c_s$ decreases with $n_B$. In all cases, we observe that the curves approach the conformal limit at asymptotically large densities. The speed of sound curves for $a=1$ exhibit the same qualitative behavior as those presented in \cite{Xia:2014zaa}. It is noteworthy that the constant speed of sound model, widely used in the literature (e.g., \cite{Alford:2013aca, Lugones:2021bkm, Ranea-Sandoval:2022izm}), corresponds to $a=2$, which is associated with a quadratic interaction among quarks.

We examined the mass-radius relationship of strange quark stars. A key feature of the curves is the significant increase in maximum gravitational mass as the effect of excluded volume is enhanced. Also, for fixed values of the parameter $a$, the excluded volume effect substantially shifts the curves towards larger radii.  The curve without excluded volume effects is significantly below the observational limit of $2~M_{\odot}$ and does not meet the observational constraints imposed by recent NICER \cite{Riley:2019yda, Miller:2019cac, Riley:2021pdl, Miller:2021qha} and LIGO/Virgo \cite{LIGOScientific:2020zkf} observations. However, by incorporating a significant fraction of excluded volume into the model, the EOS becomes stiff enough to attain the $2~M_{\odot}$ limit. Additionally, with the general increase in the radii of all objects, the other observational limits are also  fulfilled. Another interesting aspect is that models incorporating the excluded volume effect naturally satisfy observations of the recently observed object by HESS \cite{HESS2022}. This object is uniquely characterized by its very small mass and radius, which challenges theoretical models of objects containing hadronic matter \cite{Horvath:2023uwl}. As shown in Figure \ref{fig:stars}, models of strange quark matter agree more naturally with the observation of HESSJ1731.
Another noteworthy observation is GW190814, which was detected by LIGO/Virgo \cite{LIGOScientific:2020zkf}. This event involved the merger of a $23~M_{\odot}$ black hole with a compact object of  $2.6~M_{\odot}$. Should the mass of the latter be confirmed, it could either be the smallest known black hole or the most massive known neutron star. An alternative interpretation arises from our curves, suggesting that this object could be a strange quark star.   
{We have also examined the tidal deformability $\Lambda$ of strange quark stars in relation to the GW170817 event (Figure \ref{fig:tidal_def}). The model with $a=1$ more closely matches the GW170817 constraints compared to the $a=3$ model, where compatibility is limited to $\kappa=0.5$. This is because decreasing $a$ shifts the curves toward larger radii, reducing compactness and increasing tidal deformability, as illustrated in Figure \ref{fig:stars}. 
Finally, we analyzed the relationship between the baryon and gravitational masses of strange quark stars (Figure \ref{fig:baryon_mass}) and compared these findings with those derived from the commonly used SLY4 hadronic EOS. The consistent positioning of the SLY4 curve above that of the strange matter curves, suggests that the phase conversion of a metastable isolated hadronic star into a strange quark star is exothermic, in qualitative agreement with findings from previous works \cite{Benvenuto:1999uk, Berezhiani:2002ks, Bombaci:2004mt, Drago:2004vu}.
}

To conclude, it is worth comparing our results with those obtained in previous studies that used the QMDDM but employed different parameterizations of the quasiparticle mass formula.
Specifically, prior studies introduced an additional term that increases with density (see e.g. Refs. \cite{Xia:2014zaa, Li:2015ida, Issifu:2024zvq, You:2023bqx} and references therein). This inclusion enables the maximum gravitational mass of stellar configurations to fulfill the $2~M_{\odot}$ constraint. The reason for this is explained by the stability window depicted in Figure 4 of Xia et al. \cite{Xia:2014zaa}. In this figure, it becomes apparent that increasing the $\mathcal{C}$ constant in Eq. \eqref{eq:mass_eq1} requires a shift towards lower values of the $\mathcal{D}$ constant of the same equation to remain within the stability window for SQM.
By doing this, the effective mass of the quasiparticles is reduced in the density ranges typical of compact stars, thereby enhancing the EOS stiffness. However, this mass reduction effect vanishes at higher densities, as the mass formula of Eq. \eqref{eq:mass_eq1} diverges for large $n_B$.
By taking a different strategy, the effect of the mass formula used in our study ultimately yields similar outcomes.  When we consider the excluded volume effect, the interaction mass $m_I$ for quasiparticles takes the specific form  $m_I = C q^{a/3} / n_B^{a/3}$. Since $0 < q < 1$, it follows that the coefficient within the numerator of $m_I$ decreases relative to the model without excluded volume effects.  Thus, a similar effect to that obtained by \cite{Xia:2014zaa} is ultimately achieved, but with the advantage that in our case, we do not find any issues with the behavior of the mass formula at asymptotically high densities.

\vspace{6pt} 




\authorcontributions{Conceptualization, methodology, analysis and writing, G.L. and A.G.G.;  software and visualization, G.L.. All authors have read and agreed to the published version of the manuscript.}

\funding{G.L. acknowledges the financial support from  CNPq (Brazil) grant 316844/2021-7 and FAPESP (Brazil) grant 2022/02341-9. A. G. G. would like to acknowledge the financial support from CONICET under Grant PIP 22-24 11220210100150CO,  ANPCyT (Argentina) under Grant PICT20-01847, and the National University of La Plata (Argentina), Project X824.}

\dataavailability{The raw data supporting the conclusions of this article will be made available by the authors on request.}


\conflictsofinterest{The authors declare no conflict of interest.}

\abbreviations{Abbreviations}{
The following abbreviations are used in this manuscript:\\

\noindent 
\begin{tabular}{@{}ll}
QMDDM & Quark Mass Density Dependent Model\\
EOS & Equation of State\\
NS & Neutron Star\\
SQM & Strange Quark Matter \\ 
QCD & Quantum Chromodynamics 
\end{tabular}
}

\begin{adjustwidth}{-\extralength}{0cm}

\reftitle{References}


\bibliography{references}

%


\PublishersNote{}
\end{adjustwidth}

\end{document}